\documentclass[aps,pra,twocolumn,letterpaper, floatfix, superscriptaddress,amsmath,amssymb]{revtex4-2}
\usepackage{amsmath,amssymb, graphicx}
\usepackage{bm,times, color}
\usepackage[utf8]{inputenc}
\usepackage[colorlinks,breaklinks]{hyperref}
\usepackage[svgnames]{xcolor}
\hypersetup{linkcolor=DarkBlue, citecolor=DarkBlue, filecolor=black, urlcolor=DarkBlue}
\usepackage{comment, mathrsfs,slashed}
\usepackage{amsfonts}
\usepackage{bbm}
\usepackage[normalem]{ulem}
\def \p{\partial}

\def \mb{\mathbf}

\def \lan{\langle}
\def \ran{\rangle}

\def \ga{\gamma}

\begin{document}
\title{Thermoelectric properties and Wiedemann-Franz-like relations in mixed-dimensional QEDs from particle-vortex dualities}

\author{Wei-Han Hsiao}
\affiliation{Independent Researcher, Chicago, Illinois, USA}
\date{October 2021}

\begin{abstract}
We consider the thermoelectric properties of the mixed-dimensional quantum electrodynamics of the relativistic Dirac fermion and Wilson-Fisher boson. These models are self-dual, and  can form nontrivial many-body phases depending on the values of chemical potential, background magnetic field and the electromagnetic fine-structure constant. Using particle-vortex duality, we derive a variety of thermoelectric relations for strongly interacting phases with classic paradigms such as the Wiedemann-Franz law and the Mott's relation in the dual weakly interacting regimes. Besides, at the self-dual point, for the fermionic theory we find the ratio of thermal conductivity of electrical conductivity depends on the determinant of the Seebeck tensor and the phenomenological parameter Hall angle $\theta_H$. As for the bosonic theory, the dual fermion description explains how its Seebeck tensor varies depending on the dynamic regime characterized by $\theta_H$.
\end{abstract}

\maketitle

\section{Introduction}
One of the greatest puzzles in modern condensed matter physics is how strong correlations in many-body systems reshape the classic paradigms such as Drude's model and Landau's Fermi liquid theory. In the picture of the latter, up to renormalization of scales, quasiparticles essentially inherit most properties from the free particles at low energies. Phenomenological predictions have been made using these frameworks. In particular, transport properties including electric, thermoelectric and thermal responses of the quasiparticles can be conveniently explored with semiclassical kinetic equations and plenty of the results have been confirmed experimentally in solid-state systems. A renown example is the Wiedemann-Franz law, which asserts that given the temperature $T$, the ratio of thermal conductivity $\kappa$ to electric conductivity $\sigma$ equals a universal constant, the Lorenz number $\mathcal L_0$,
\begin{align}
\label{WF_law}\frac{\kappa}{\sigma T} =\mathcal L_0 =  \frac{\pi^2}{3}.
\end{align}
In the past two decades, enormous works have discovered violations of {\it laws} like~\eqref{WF_law} and identified them as signatures of physics beyond Landau's spell. Examples occur frequently in systems having separate electric and thermal transports contributors such as fractional quantum Hall states, vortex metals, and the vicinity of heavy-fermion quantum criticality \cite{PhysRevB.55.15832, PhysRevLett.95.077002, PhysRevLett.102.156404}. Rather interestingly, it is also demonstrated that similar violations could happen to arbitrary an extent \cite{PhysRevLett.115.056603, PhysRevB.97.245128, PhysRevB.99.085104} even within the framework of Fermi liquid theory, where the driving force behind the violations is nontrivial inelastic scattering processes.

Having acknowledged the breakdown of relations like Eq.~\eqref{WF_law} does not sufficiently imply a new ballpark for many-body physics, we would like to investigate the properties of thermoelectric transports from an alternative perspective by exploring new thermoelectric relations that may reveal new insights for strongly coupled-many-body systems. In this regard, dualities between quantum field theories offer a natural platform because of their capabilities of bridging different phases of matters. For instance, a Wiedemann-Franz-like relation has been derived for (2+1)-dimensional conformal field theories (CFTs) with holographic duals \cite{Melnikov2021}. In a similar spirit, we aim to explore thermoelectric properties for some dualities in the context of (2+1)-dimensional matters coupling to (3+1)-dimensional electromagnetism. More precisely, in terms of action, we consider the mixed-dimensional quantum electrodynamics, dubbed QED$_{4,3}$, of the form 
\begin{align}
S = S_{\rm{matter}}[\{\phi\}, A] - \frac{1}{4e^2}\int d^4x\, F_{\mu\nu}F^{\mu\nu},
\end{align}
where $\{\phi\}$ is the collection of fields living in pure (2+1) dimensions. Such a collection could consist of matter fields or emergent dynamical gauge fields responsible for flux attachment. The simplest examples of $S_{\rm{matter}}$ and also the foci of this work are the two-component Dirac fermion \footnote{The parity anomaly would not concern us since we have coupled the Dirac cone to a (3+1)-dimensional field \cite{PhysRevB.88.085104}.}
\begin{subequations}
\begin{align}
\label{Dirac_fermion}\mathscr L_f = \bar{\psi}i\slashed{D}_A\psi,
\end{align}
and the Wilson-Fisher boson
\begin{align}
\label{WF_boson}\mathscr L_{b} = |D_A\phi|^2-|\phi|^4
\end{align}
where $(D_A)_{\mu} = \p_{\mu} - iA_{\mu}$ denotes the conventional covariant derivative in the flat spacetime. 
\end{subequations} 
When coupled to the (3+1)-dimensional Maxwell term, both Eqs.~\eqref{Dirac_fermion} and~\eqref{WF_boson} acquire the structure of self-duality \cite{PhysRevB.96.075127, PhysRevB.100.235150}, that is to say, the full partition function $Z[S[e]] = Z[S[\tilde{e}]]$ with $\tilde{e} \sim 1/e$. This result relies upon that both matter theories admit the vortex-dual descriptions via the celebrated particle-vortex duality \cite{PESKIN1978122, PhysRevLett.47.1556, SEIBERG2016395, PhysRevX.6.031043}. Remarkably, at the self-dual value of coupling constant $e = \tilde e$, one can make quantitative inferences about the coefficients of electric, thermoelectric, and thermal transports with the background electromagnetic field specified \cite{PhysRevB.96.075127, PhysRevB.100.235150}. These relations will be exploited further in this work. In addition, the structure of strong-weak duality $e\leftrightarrow \tilde{e}$ opens up a route for us to explore strongly interacting phase $e\gg 1$ via the weakly interacting counterpart $\tilde{e}\ll1$. In the IR this approximation is much more controlled compared to QED3 because of the marginal nature of the (3+1)-dimensional charge. In particular, for the spinor QED${}_{4,3}$, it has been argued that the beta function $\beta(e^2)$ vanishes at charge neutrality, and $e^2$ does not run \cite{PhysRevD.86.025005, PhysRevD.97.074004, PhysRevD.99.045017}, helping justify perturbative treatments given a fine structure constant of appropriate size. 

Regarding the thesis of this work, for model~\eqref{Dirac_fermion}, we shall leverage some properties of the QED${}_{4,3}$ in perturbative regime and at the self-dual point to derive thermoelectric properties in strongly interacting regimes and at the self-dual point itself. Specifically, we shall map some non-trivial phases to those which permit free quasiparticle descriptions such as a Fermi liquid or an integer quantum Hall state. By virtue of duality, we can derive a bound of deviation from Eq.~\eqref{WF_law} for single Dirac cone at charge neutrality with infinite strength of electromagnetic interaction. We can also show, in the particle-hole symmetric phase in the zeroth Landau level, the ratio of the longitudinal components of open-circuit thermal conductivity $\bar{\kappa}$  to resistivity $\rho$ assumes a constant value in the strongly interacting regime $e\gg 1$. As for model~\eqref{WF_boson}, we will consider its fermionized description at the self-dual point, where the composite fermions form a Fermi surface with purely time-reversal even dynamics, and derive some nonperturbative thermoelectric relations in terms of the phenomenological parameter Hall angle $\theta_H$. 

The rest of the paper is structured as follows. In Sec. \ref{duality_review} we review the essential results from mixed-dimensional particle-vortex duality that will be utilized. These include the mappings between different phases given the values of particle-density $n$ and background magnetic field $B$, the transformation rules of electric, thermoelectric and thermal conductivities under duality, the effect of mixed-dimensional Maxwell action, and the role of the coupling constant $e^2$. The main results of the work will be presented in Sec. \ref{main_results}. We will look into several phases of matter in strongly interacting regime $e^2\gg 1$ and the self-dual points of the spinor and scalar QED${}_{4,3}$s. By mapping them into weakly interacting phases or constraining them by the structure of self-duality, quantitative relations or identities about their thermoelectric transports are derived. Then we will wrap up the work in \ref{summary} with some comments and open directions.

\section{Dualities, relations between phases and transport coefficients}\label{duality_review}

This section shall refresh how transport coefficients are related under particle-vortex dualities. We start with dualizing Eq.~\eqref{Dirac_fermion} by the fermionic particle-vortex duality \cite{PhysRevX.5.031027, PhysRevX.5.041031, PhysRevB.93.245151}
 \begin{align}
\label{fermion_particle_vortex} \mathscr L_f(A) \leftrightarrow \bar{\chi}i\slashed{D}_a\chi + \frac{1}{4\pi}a\, dA
 \end{align}
with the standard shorthanded notation $a\, dA = \epsilon^{\mu\nu\lambda}a_{\mu}\p_{\nu}A_{\lambda}$. The lower-cased $a_{\mu}$ denotes the statistical gauge field responsible for relativistic flux attachments. We shall denote the corresponding field strengths $da$ using lower cases $(b, \mb e)$ in the rest of the section. 

The mappings between operators can be derived by varying gauge fields coupling to the conserved currents. For example, 
 \begin{subequations}
 \begin{align}
\label{current_psi} J^{\mu}_{\psi} = \bar{\psi}\ga^{\mu}\psi = \frac{1}{4\pi}\epsilon^{\mu\nu\lambda}\p_{\nu}a_{\lambda}.
 \end{align}
 The current $J^{\mu}_{\chi}$ is the on-shell condition for $a_{\mu}$.
 \begin{align}
\label{current_chi} J^{\mu}_{\chi} = \bar{\chi}\ga^{\mu}\chi = -\frac{1}{4\pi}\epsilon^{\mu\nu\lambda}\p_{\nu}A_{\lambda}.
 \end{align}
 \end{subequations}
 Equations~\eqref{current_psi} and~\eqref{current_chi} summarize the essence of particle-vortex dualities: The charge currents of the particles manifest themselves in terms of electromagnetism in the dual description. It is also worth clarifying that while the field $\psi$ is charged under $A_{\mu}$, the field $\chi$ is not. Rather, it is charged under the emergent field $a_{\mu}$, which is dual to the densities of $\psi$. 
The quantum phases of the theories are characterized by the ratio of particle number density to that of the magnetic fluxes $\nu_{\psi} = 2\pi n_{\psi}/B$ and $\nu_{\chi} = 2\pi n_{\chi}/b$. The zeroth components \footnote{To avoid confusion, in this work we adopt the convention $A_{\mu} = (A^0, -\mb A)$ for U(1) gauge fields. Therefore, $\epsilon^{ij}\p_iA_j = -B$ and $-\p_0A_i - \p_i A_0 = E_i$. It is straightforward to verify this convention consistently yields all electromagnetism.} of Eq.~\eqref{current_psi} and Eq.~\eqref{current_chi} together yield 
\begin{subequations}
\begin{align}
\label{fermion_phase_mapping} 2\nu_{\psi} = -\frac{1}{2\nu_{\chi}}.
\end{align}
The spatial components entail the correspondences of charge currents and dual electric fields.
\begin{align}
\label{current_psi_in_e} & J^i_{\psi} = -\frac{1}{4\pi}\epsilon^{ij}e_j\\
\label{current_chi_in_E}& J^i_{\chi} = \frac{1}{4\pi}\epsilon^{ij}E_j.
\end{align}
 \end{subequations}
To relate the transport coefficients on two sides of the duality, we consider the formal definition of conductivities.
\begin{subequations}
\begin{align}
\label{current_psi_linear}& \mb J_{\psi} = \sigma_{\psi}\mb E + \alpha_{\psi}(-\nabla T)\\
\label{current_chi_linear}& \mb J_{\chi} = \sigma_{\chi}\mb e + \alpha_{\chi}(-\nabla T),
\end{align}
\end{subequations}
where the temperature $T$ is assumed to be invariant under duality. As we pointed out in the above, $\psi$ and $\chi$ are driven by different electric fields since they are charged under distinct gauge fields. Using Eq.~\eqref{current_psi_in_e} to write $\mb J$ in terms of $\mb e$, $-\frac{1}{4\pi}\epsilon\mb e = \sigma\mb E + \alpha(-\nabla T).$ Next, we invert Eq.~\eqref{current_chi_linear} and rewrite $\mb J_{\chi}$ similarly using Eq.~\eqref{current_chi_in_E}. This sequence of operations gives 
\begin{align*}
-\frac{1}{4\pi}\epsilon\sigma_{\chi}^{-1}\left( \frac{1}{4\pi}\epsilon\mb E + \alpha_{\chi}\nabla T\right) = \sigma_{\psi}\mb E + \alpha(-\nabla T).
\end{align*}
Since $\mb E$ and $\nabla T$ vary independently, in order for the above equation to be consistent, we conclude
\begin{align}
\label{electric_relation} & \sigma_{\psi} = \frac{1}{(4\pi)^2}\epsilon^{-1}\sigma_{\chi}^{-1}\epsilon = \frac{1}{(4\pi)^2}\epsilon^{-1}\rho_{\chi}\epsilon\\
\label{thermoelectric_relation}& \alpha_{\psi} = \frac{1}{4\pi}\epsilon\sigma_{\chi}^{-1}\alpha_{\chi}.
\end{align}
Note that Eq.~\eqref{thermoelectric_relation} can be further elucidated by introducing the Seebeck tensor defined by $\mb E = S\nabla T$ in the absence of electric current. By Eq.~\eqref{current_chi_linear}, $S_{\chi} = \sigma^{-1}_{\chi}\alpha_{\chi}$, and thus equivalently 
\begin{align}
\alpha_{\psi} = \frac{1}{4\pi}\epsilon S_{\chi}.
\end{align}
Next we consider the linear response of heat current 
\begin{align}
\mb q = T\alpha_{\psi}\mb E - \bar{\kappa}_{\psi}\nabla T.
\end{align}
$\bar{\kappa}$ is understood as the thermal conductivity in the absence of electric field. Unlike the particle density current, the physical heat current does not transform under duality, and simply has two descriptions in terms of $\psi$ and $\chi$ fields. Consequently, we identify 
\begin{align*}
T\alpha_{\psi}\mb E - \bar{\kappa}_{\psi}\nabla T = T\alpha_{\chi}\mb e - \bar{\kappa}_{\chi} \nabla T.
\end{align*}
Substituting the inverse of Eq.~\eqref{current_chi_linear} for $\mb e$ again, the consistency condition implies 
\begin{align}
\label{thermal_relation}\bar{\kappa}_{\psi} = \bar{\kappa}_{\chi} - T \alpha_{\chi} \sigma^{-1}_{\chi}\alpha_{\chi} \equiv \kappa_{\chi}.
\end{align}
$\kappa$ is the thermal conductivity in the absence of current, or the open circuit thermal conductivity. This correspondence is not surprising because electric currents and fields swap the roles under particle-vortex duality and so must $\kappa$ and $\bar{\kappa}$. Note that the above derivations could be directly applied to bosonic particle-vortex duality, which states 
\begin{align}
\label{boson_particle_vortex}\mathscr L_b(A)\leftrightarrow |D_a\Phi|^2 - |\Phi|^4 +\frac{1}{2\pi}a\, dA.
\end{align}
The results are almost the same except the $(4\pi)s$ in Eqs.~\eqref{electric_relation} and~\eqref{thermoelectric_relation} are replaced with $(2\pi)$. Nevertheless, we wish to focus on the fermionized description of Eq.~\eqref{WF_boson}, which is given by relativistic flux attachment 
\begin{align}
\label{boson_fermi_duality}\mathscr L_b(A)\leftrightarrow i\bar{\Psi}\slashed D_a\Psi -\frac{1}{8\pi} a\, da -\frac{1}{2\pi}a\, dA -\frac{1}{4\pi}A\, dA.
\end{align}
The mappings between field operators can be derived again by varying $a$ and $A$. 
\begin{subequations}
\begin{align}
\label{current_phi}& J^{\mu}_{\phi} = -\frac{1}{2\pi} \star da - \frac{1}{2\pi} \star dA\\
\label{current_Psi}& J^{\mu}_{\Psi} = \frac{1}{4\pi} \star da + \frac{1}{2\pi} \star dA.
\end{align}
\end{subequations}
The star $\star$ denotes the standard Hodge dual. The zeroth components of the above two equations give rise to 
\begin{align}
\label{boson_phase_mapping}(\nu_{\phi} - 1)\left(\nu_{\Psi} + \frac{1}{2}\right) = -1.
\end{align}
An immediate consequence is when $\nu_{\phi} = 1$, the fermion $\Psi$ sees an average magnetic field $\lan b\ran = -\lan (\star da)^0\ran = 0$ and thus forms a Fermi liquid in the mean-field approximation. By considering the spatial components of Eqs.~\eqref{current_phi} and~\eqref{current_Psi} and the linear response identities parallel to Eqs.~\eqref{current_psi_linear} and~\eqref{current_chi_linear}, one can establish relations between fermionic and bosonic transport coefficients by the same token as one presented in the earlier part of the section.
\begin{subequations}
\begin{align}
\label{electric_relation_bf}& \epsilon\left(\sigma_{\phi} - \frac{\epsilon}{2\pi}\right)\epsilon^{-1}\left(\sigma_{\Psi} + \frac{\epsilon}{4\pi}\right) = \frac{1}{(2\pi)^2}\\
\label{thermoelectric_relation_bf}& \alpha_{\phi} = -\frac{1}{2\pi}\epsilon\left(\sigma_{\Psi} + \frac{1}{4\pi}\epsilon\right)^{-1}\alpha_{\Psi}\\
\label{thermal_relation_bf}& \bar{\kappa}_{\phi} = \bar{\kappa}_{\Psi} - T \alpha_{\Psi} \left(\sigma_{\Psi} + \frac{1}{4\pi}\epsilon\right)^{-1}\alpha_{\Psi}.
\end{align}
\end{subequations}
Note that Eqs.~\eqref{electric_relation},~\eqref{thermoelectric_relation},~\eqref{thermal_relation},~\eqref{electric_relation_bf},~\eqref{thermoelectric_relation_bf}, and~\eqref{thermal_relation_bf} are exact and valid given any values of momentum $(\omega, \mb q)$. In 2-spatial dimensions, rotational symmetry decomposes a matrix $w^{ij}(\omega, \mb q) =w_{L}q^iq^j + w_T(q^iq^j -\delta^{ij}\mb q^2) + w_H\epsilon^{ij}$, where the form factors $w_L, w_T$, and $w_H$ are functions of $\mb q^2$ and $\omega$. In the long-wavelength limit $\mb q\to 0$, the decomposition reduces to $w^{ij} = w_{xx}\delta^{ij} + w_{xy}\epsilon^{ij}$, and some identities can be simplified. For instance, Eq.~\eqref{electric_relation} reduces to $\sigma_{\psi}\sigma_{\chi} = \frac{1}{(4\pi)^2}.$ In the rest of the paper, we shall assume the long-wavelength limit and concentrate on the {\it optical responses.}

We would like to conclude this section by mentioning that Eqs.~\eqref{electric_relation},~\eqref{thermoelectric_relation}, and~\eqref{thermal_relation} are ubiquitous in models admitting dual vortex descriptions, which range from 3-dimensional XY model to holomorphic dualities where the bulk gravity theory acquires electromagnetic duality \cite{doi:10.1142/S0217979290000206, PhysRevB.63.155309, PhysRevB.76.144502, Donos2017, PhysRevD.76.106012}. The addition of the mixed-dimensional Maxwell term endows these relations with more precise implications. To be explicit, turning on $e^2$ and the Maxwell term in both sides of Eq.~\eqref{fermion_particle_vortex}, it can be shown by integrating over out-of-plane degrees of freedom that the following theory 
\begin{align}
\label{sd_QED}S[e] = \int d^3x\, \mathscr L_f(A)-\frac{1}{4e^2}\int d^4x\, F_{\mu\nu}F^{\mu\nu}
\end{align}
is dual to the same action $S[\tilde{e}]$ with a different coupling constant
\begin{align}
\label{coupling_mapping}\tilde{e} = \frac{8\pi}{e}.
\end{align}
In short, the theory~\eqref{sd_QED} is self-dual and exhibits the structure of strong-weak duality. The assertion is true when applied to Eq.~\eqref{boson_particle_vortex} with the replacement $\tilde e' = (4\pi)/e$. As we will elaborate in the following section, Eqs.~\eqref{fermion_phase_mapping},~\eqref{boson_phase_mapping}, and~\eqref{coupling_mapping} offer us tools to describe some nontrivial strongly interacting phases in terms of their weakly interacting dual counterparts controlled by the marginal value $e^2$ and explore properties at the self-dual point $e = \tilde e$.

\section{Thermoelectric properties by dualities}\label{main_results}
This section will be denoted to various thermoelectric implications for the mixed-dimensional Dirac fermion and Wilson-Fisher boson models using the relations between transport coefficients derived in the last section. We shall first investigate various strongly interacting phases of the model~\eqref{sd_QED} at zero density and magnetic field. Next we turn on chemical potential and magnetic field, looking into phases characterized by the filling factor and the electromagnetic coupling $(\nu, e^2)$. These phases are explored using their weakly interacting duals. Then we will discuss aspects of the Wilson-Fisher boson at the self-dual point from the composite fermion picture~\eqref{boson_fermi_duality}. At the bosonic self-dual point, characterized by $(\nu_{\phi}, e^2)\to (1, 4\pi)$, the dual fermionic vortices form an time-reversal even phase with vanishing Hall components for all conductivity tensors, which provides us an algebraically feasible route to retrieve knowledge about the dual bosons. 
\subsection{$n_{\psi} = 0, B = 0, e^2\to\infty$}
In this configuration, both $\psi$ and $\chi$ sit at the charge neutrality, where typical Fermi liquid theory breaks down because $T/T_F \gg 1$. The dynamics of $\psi$ is expected to be affected by the $A_{\mu}$ fields, whereas $\chi$ forms a nearly free Dirac cone. In the absence of the magnetic field, the  coefficients of optical responses are not only isotropic but time-reversal even, i.e., that $w^{ij}(\omega) = w(\omega)\delta^{ij}.$ This symmetry consideration and Eq.~\eqref{electric_relation} imply $\sigma_{\psi}(e=\infty)= 1/[(4\pi)^2\sigma_{\chi}(\tilde e =0)] = 1/\pi^2$. The thermoelectric response $\alpha_{\psi}\propto \epsilon \sigma^{-1}_{\chi}\alpha_{\chi}$ can only be zero for the sake of self-consistency, which, together with Eq.~\eqref{thermal_relation}, in turn implies $\bar{\kappa}_{\psi}(e) = \bar{\kappa}_{\chi}(\tilde e)$ and $\kappa_{\psi}(e)= \kappa_{\chi}(\tilde{e})$. Particle-vortex duality alone fixes the value of thermal conductivity at charge neutrality. Let us look at the ratio indicating the deviation from the Wiedemann-Franz law.
\begin{align*}
r_{\chi}(\tilde{e}) = \frac{1}{\mathcal L_0}\frac{\kappa_{\chi}(\tilde{e})}{T\sigma_{\chi}(\tilde{e})}.
\end{align*}
While it is well known that $r$ could be as large as $\mathcal O(10)$ in the vicinity of a strongly interacting quantum critical point in graphenelike systems \cite{Crossno1058, PhysRevB.93.075426}, it is shown that for free Dirac fermions $r$ also exceeds unity and is sandwiched in between two and three \cite{PhysRevB.67.144509, ma14112704}. Using $\kappa_{\psi}(e)= \kappa_{\chi}(\tilde{e})$ and $r_{\psi}(\infty) = r_{\chi}(0)\frac{\sigma_{\chi}(0)}{\sigma_{\psi}(\infty)}$, we could estimate a bound for deviation of the Wiedemann-Franz law for the spinor QED${}_{4,3}$ in the strongly interacting limit. 
\begin{align}
 \label{lorenz_bound}1.23< 2\times \frac{\sigma_{\chi}(0)}{\sigma_{\psi}(\infty)} <r_{\psi}(\infty) < 3\times\frac{\sigma_{\chi}(0)}{\sigma_{\psi}(\infty)}<1.86.
\end{align}
This bound nontrivially quantifies not only the amount the relativistic QED${}_{4,3}$ breaks the Wiedemann-Franz law in the limit of infinite strength of coupling but also its discrepancy from the non-relativistic graphene systems near the inelastic-scattering dominating Dirac points.
\subsection{$(\nu_{\psi}, e^2)\to (0, \infty), B\neq 0$}
In this phase, the $\psi$ particle forms a charge-conjugation symmetric fractional quantum Hall state in the zeroth Landau level because of the magnetic field and strong fluctuations of mixed-dimensional electromagnetism. This is the relativistic counterpart of the half-filled Landau level in the contex of conventional quantum Hall effect. In the dual description, the fermionic vortex $\chi$ forms a Fermi liquid because of the absence of average magnetic field $|b|\propto n_{\psi} = 0$. The depth of the Fermi sea at $T = 0$ is given by $p_F = \sqrt{|B|}$ owing to $n_{\chi} \propto B/(2\pi)$ and $p_F \propto n^{1/2}_{\chi}$. The Fermi surface couples to a perturbative electromagnetic field controlled by strength $\tilde e \to 0$. On one hand, in the semi-classical limit, the ratio $(\kappa_{\chi})_{xx}/[T(\sigma_{\chi})_{xx}]$ is given by the Wiedemann-Franz law~\eqref{WF_law}. On the other hand, by Eqs.~\eqref{electric_relation} and~\eqref{thermal_relation}, entry-wise this ratio is dual to 
\begin{align}
\frac{(\kappa_{\chi})_{xx}}{T(\sigma_{\chi})_{xx}} = (4\pi)^2\frac{(\bar{\kappa}_{\psi})_{xx}}{T(\rho_{\psi})_{xx}}.
\end{align}
Consequently, in the strongly interacting limit, the fractional quantum Hall state respects a Wiedemann-Franz-like relation \footnote{Note that the dimensionless $\sigma$ measures the coefficient of $e^2/\hbar$ for the empirical conductivity. Therefore, the $\rho$ in this equation should also be understood as the coefficient of $\hbar/e^2$ for the empirical resistivity in the corresponding system.}
\begin{align}
\label{WF_generalized}\frac{(\bar{\kappa}_{\psi})_{xx}}{T(\rho_{\psi})_{xx}} = \frac{1}{48}.
\end{align}
We could also consider the Mott's law for the Fermi liquid. 
\begin{align}
\alpha_{\chi} = \frac{\pi^2}{3}T\frac{\p\sigma_{\chi}}{\p\mu_{\chi}}.
\end{align}
By Eq.~\eqref{electric_relation},
\begin{align*}
\frac{\p\sigma_{\chi}}{\p\mu_{\chi}} = \frac{1}{(4\pi)^2}\frac{\p\rho_{\psi}}{\p\mu_{\chi}}.
\end{align*}
For a relativistic Fermi liquid at low temperature, the leading order $\mu_{\chi}$ is given by $p_F = \sqrt{|B|}$. Together with~\eqref{thermoelectric_relation}, we can deduce 
\begin{subequations}
\begin{align}
\epsilon\frac{\p\rho_{\psi}}{\p|B|} = \frac{6}{\pi |B|T}\sigma^{-1}_{\psi}\alpha_{\psi} = \frac{6}{\pi |B|T}S_{\psi},
\end{align}
or equivalently 
\begin{align}
\alpha_{\psi} = \frac{\pi T|B|}{6}\epsilon\frac{\p}{\p |B|} \ln \rho_{\psi}.
\end{align}
These identities express thermoelectric responses in terms of derivatives of electric responses with respect to the magnitude of the magnetic field in the leading order of temperature expansion. We could also straightforwardly infer the vanishing of diagonal entries of the Seebeck tensor $(S_{\psi})_{xx}\propto (\alpha_{\chi})_{xy} = 0$ because there is no time-reversal symmetry breaking mechanism in the $\chi$ description. 

Recall that the phase of $\psi$ field is the relativistic and mixed-dimensional analog of the $\nu_{\rm{NR}} = 1/2$ problem \cite{PhysRevX.5.031027}. As a consequence, Eq.~\eqref{WF_generalized} is also reminiscent of the implication that the $\nu_{\rm NR} = \frac{1}{2}$ state, or a general two-dimensional quantum fluid with a vortex description, violates the conventional Wiedemann-Franz law \cite{PhysRevLett.95.077002, PhysRevB.93.085110}. The underlying physical reason is that the Fermi liquid is formed by fermionic vortices and therefore it is the ratio of the vortices' electric conductivity to thermal conductivity that abides by the Wiedemann-Franz law~\eqref{WF_law}. Although the observable heat current remains equivalent under different descriptions, the charge currents and electric fields swap under the particle-vortex duality. As such, there is a separation of charge and heat conducting degrees of freedom in the $\psi$ description, breaking the canonical paradigm. 
\end{subequations}
Nevertheless, despite similarly utilizing a Fermi surface formed by Dirac-type composite fermions, the thermoelectric inference by the mixed-dimensional QED could be rather different than ones predicted by the purely (2+1)-dimensional quantum Hall models \cite{PhysRevX.6.031026} because the necessity of an additional $\frac{1}{2}\frac{1}{4\pi} A\, dA$ term in the action. For instance the Seebeck coefficient could be nonvanishing because the thermoelectric contribution from this shift in the electric response. Note that, nonetheless, $\alpha_{xx} = 0$ in both examples. 

\subsection{$(\nu_{\psi}, e^2)\to (-\frac{1}{2}, \infty)$}
In this phase, because the relativistic Landau levels are labeled by half-integers, $\psi$ particles fill the Landau levels right beneath the zeroth Landau level, whereas $\chi$ particles fill the zeroth Landau level with small strength of interaction, or a simple integer quantum Hall state. Again we have a relation Eq.~\eqref{WF_generalized} for the off-diagonal Hall response. In the nearly ballistic limit $T\ll \omega \ll n^{1/2}$, $(\sigma_{\chi})_{xy} = \frac{1}{2}\frac{1}{2\pi}$.  
\begin{subequations}
\begin{align}
\label{thermal_hall_one}(\bar{\kappa}_{\psi})_{xy} = \frac{T}{48}(4\pi)^2(\sigma_{\chi})_{xy}= \frac{\pi}{12}T.
\end{align}
We could extend this example to general integer quantum Hall states of $\chi$ labeled by $\nu_{\chi} = \pm(N+1/2)$. They are dual to strongly interacting fractional quantum Hall phases at filling fractions $\mp \frac{1}{2(2N+1)}$, or the relativistic version of the Jain sequence. At any integral value of $N$,~\eqref{thermal_hall_one} naturally generalizes to
\begin{align}
\label{thermal_hall_two}(\bar{\kappa}_{\psi})_{xy} = \frac{\pi}{6}\left(N+\frac{1}{2}\right)T.
\end{align}
\end{subequations}
We have not understood how the large $N$ limit should be carried out in order to establish a smooth interpolation between the result and a Fermi liquid phase. A reasonable resolution would be performing the $N$ re-summation in the random phase approximation (RPA) before gradient expansion in $(\omega, \mb q)$ by the same token as the treatment in Ref. \onlinecite{PhysRevB.95.085151}.

\subsection{$(\nu, e^2)\to (\frac{1}{2}, 8\pi)$}
This corresponds the self-dual point of the spinor mixed-dimensional particle-vortex duality Eq.~\eqref{sd_QED}. $\psi$ and $\chi$ particles experience the same strength of electromagnetic interaction, and their phases are related by the time-reversal transformation. In terms of the tensors of linear response, 
\begin{subequations}
\begin{align}
\label{electric_sd}& \sigma_{\chi} = \mathcal T\sigma_{\psi}\mathcal T^{-1} = \sigma_{\psi}^T\\
\label{thermoelectric_sd}& \alpha_{\chi} = \mathcal T\alpha_{\psi}\mathcal T^{-1} = \alpha_{\psi}^T\\
\label{thermal_sd}& \kappa_{\chi} = \mathcal T\kappa_{\psi}\mathcal T^{-1} = \kappa_{\psi}^T.
\end{align}
\end{subequations} 
Equations~\eqref{electric_relation} and~\eqref{electric_sd} imply the semicircle law for the conductivity tensor 
\begin{align}
\sigma_{xx}^2 +\sigma_{xy}^2 = \frac{1}{(4\pi)^2}
\end{align}
We suppress the $\psi$ label since at this point $\psi$ and $\chi$ are essentially time-reversal counterparts. These entries can then be parametrized by the phenomenological parameter Hall angle $\theta_H$: $\sigma_{xx} = \frac{1}{4\pi}\cos\theta_H$ and $\sigma_{xy} = \frac{1}{4\pi}\sin\theta_H$. Introducing $\theta_H$ to Eqs.~\eqref{thermoelectric_relation} and~\eqref{thermoelectric_sd}, the ratio 
 \begin{subequations}
 \begin{align}
\frac{\alpha_{xy}}{\alpha_{xx}} = \tan\left(\frac{\pi}{4} +\frac{\theta_H}{2}\right).
 \end{align}
 Alternatively the ratio of the off-diagonal and diagonal components of the Seebeck tensor reads 
 \begin{align}
\frac{S_{xy}}{S_{xx}} = \tan\left(\frac{\pi}{4}-\frac{\theta_H}{2}\right).
\end{align}
\end{subequations}
To proceed, we can use Eqn~\eqref{thermal_sd} $\bar{\kappa}_{\psi} - \bar{\kappa}_{\chi} = \bar{\kappa}_{\psi}-\bar{\kappa}^T_{\psi} = -T\alpha_{\chi}S_{\chi}$. The new implication of the above transport quantities is that we can explore a Wiedemann-Franz-like relation at the self-dual point:
\begin{subequations}
\begin{align}
\label{WF_sd_one} \frac{\bar{\kappa}_{xy}}{T\sigma_{xy}} =\frac{{\kappa}_{yx}}{T\sigma_{xy}} = \frac{1}{2\sin\theta_H}(S_{xx}^2 + S_{xy}^2).
\end{align}
We can look at different dynamic regimes by tuning the parameter $\theta_H$. In particular, in the quantum Hall regime $\theta_H \to \pi/2$, $\alpha_{xy}/\alpha_{xx} \to \infty$ and $S_{xy}/S_{xx}\to 0$. The ratio~\eqref{WF_sd_one} converges to
\begin{align}
\label{WF_sd_two}\frac{\bar{\kappa}_{xy}}{T\sigma_{xy}} = 8\pi^2\alpha^2_{xy} = \frac{1}{2}S^2_{xx}.
\end{align}
\end{subequations}
The generalized {\it Lorenz number} is given by the square of the diagonal component of the Seebeck tensor, or that of the Hall component of the thermoelectric conductivity. We note that unlike Eqs.~\eqref{thermal_hall_one} and~\eqref{thermal_hall_two}, which still require some knowledge about the weakly interacting phases on one side of the duality, Eqs.~\eqref{WF_sd_one} and~\eqref{WF_sd_two} are exact to all orders in the coupling constant $e^2$ at a nonperturbative value $e^2 = 8\pi$ and emerges solely from self-duality of the theory.
\subsection{$(\nu_{\phi}, e^2)\to (1, 4\pi)$} 
Finally, we look at the self-dual point of the mixed-dimensional Wilson-Fisher theory~\eqref{boson_particle_vortex}. It exhibits universal transport properties similar to Eqs.~\eqref{electric_sd},~\eqref{thermoelectric_sd}, and~\eqref{thermal_sd}. Physically, this configuration is populated with equal density of boson and magnetic flux tubes. This is reminiscent of the field-induced superconductor-insulator transition (SIT) \cite{PhysRevB.93.205116, PhysRevB.95.045118} and the $\nu = 1$ bosonic fractional quantum Hall state \cite{PASQUIER1998719, PhysRevB.58.16262}. Naively we could repeat all the arguments in the above to deduce inferences for some bosonic phases strongly interacting with the electromagnetic interaction. However, the nature of the electric and thermal transport properties at the Mott-superfluid transition is not as transparent and simple as the fermionic counterpart. As such, we attempt to describe this phase in terms of the mixed-dimensional extension of Eq.~\eqref{boson_fermi_duality}, which implies the fermion field $\Psi$ forms a Fermi surface and interacts with emergent gauge field $a_{\mu}$, which is charged under $A_{\mu}$.
 
In this example the bosonic electric transport abides by the semicircle law $(\sigma_{\phi})^2_{xx} + (\sigma_{\phi})^2_{xy}= \frac{1}{(2\pi)^2}.$ We can parametrize the dynamic regime similarly with the Hall angle $\theta_H$. Using Eq.~\eqref{electric_relation_bf}, it can be shown the fermion field has time-reversal even electric conductivities.
\begin{align}
(\sigma_{\Psi})_{ij} = \frac{1}{4\pi}\tan\left(\frac{\theta_H}{2}+\frac{\pi}{4}\right) \delta_{ij}.
\end{align}
Using this parametrization, thermoelectric properties can be further elucidated from the fermionic aspect. Taking the Seebeck coefficient at the self-dual point for example, 
\begin{align}
S_{\phi} =\sigma_{\phi}^{-1}\alpha_{\phi} =  -\frac{1}{2\pi}\sigma^{-1}_{\phi}\epsilon\left(\sigma_{\Psi} + \frac{1}{4\pi}\epsilon\right)^{-1}\alpha_{\Psi}.
\end{align}
In the quantum Hall regime for $\phi$: $\theta_H\to\pi/2$, the $\Psi$ field approaches the ballistic limit and the anomalous Hall contribution becomes negligible $\left(\sigma_{\Psi} + \epsilon/(4\pi)\right)^{-1}\approx \sigma_{\Psi}^{-1}$. Consequently, 
\begin{subequations}
\begin{align}
\theta_H \to \frac{\pi}{2}: \label{S_phi_in_psi_one}(S_{\phi})_{ij} \approx -\frac{\alpha_{\Psi}}{\sigma_{\Psi}}\delta_{ij}.
\end{align}
For the self-dual boson, its Seebeck tensor also becomes diagonal because it is dual directly to the fermionic Seebeck tensor, which has to be diagonal since there is no breakdown of time-reversal symmetry. In the opposite limit $\theta_H \to 0$, $\sigma_{\Psi}$ saturates to the same order of the anomalous Hall conductivity $\left(\sigma_{\Psi} + \epsilon/(4\pi)\right)^{-1}\approx 2\pi(1-\epsilon).$
\begin{align}
\theta_H \to 0: \label{S_phi_in_psi_two}S_{\phi} = -\frac{\alpha_{\Psi}}{2\pi}(1+\epsilon)
\end{align}
\end{subequations}
We arrive at $(S_{\phi})_{xx}= (S_{\phi})_{xy}$. 

Ideally, at the mean-field level $\alpha_{\Psi}\propto T\frac{\p\sigma_{\Psi}}{\p\mu}$, $\alpha_{\Psi}/\sigma_{\Psi}$ of the Fermi liquid could give an estimate of temperature dependence of $S_{\phi}$ and $\kappa_{\phi}$ when applied to Eqs.~\eqref{S_phi_in_psi_one} and~\eqref{S_phi_in_psi_two}. In fact, for this fermionized model, it can be shown that an isotropic, time-reversal even and free fermion approximation could fulfill the self-dual condition. This algebraic fact explains why similar conclusions can be derived near the SIT with a similar model in the RPA \cite{PhysRevB.95.045118}. This, in our context, would unfortunately only be a very rough approximation since $e^2 = 4\pi$ by no means corresponds to a weakly interacting phase. In fact, owing to fluctuating $a_{\mu}$, the effective coupling $e_{\rm{eff}}^2$ coupled to $\Psi$ is given by 
\begin{align}
e_{\rm{eff}}^2 = e^2\left[ 1+ (4\pi/e^2)^2\right]\geq 8\pi.
\end{align}
The $\Psi$ fermions would always interact furiously with the gauge fluctuations. The bosonic self-dual point $e^2 = 4\pi$ already offers the most optimistic strength and in addition it is the only value at which the time-reversal-symmetric assumption is validate. Therefore, we shall not attempt to make further inferences using the duality technique.
\section{Conclusion}\label{summary}
We considered the thermoelectric properties of the relativistic Dirac and Wilson-Fisher mixed-dimensional quantum electrodynamics. Depending on the values of chemical potential, background magnetic field, and the fine structure constant, these models depict various non-trivial many-body phases on a thin film such as the Dirac liquid at the vicinity of the Dirac point and the relativistic versions of the half-filled Landau level problem for both fermions and bosons. By exploiting the weakly interacting vortex descriptions and the properties of self-duality of these models and phases, a variety of intriguing thermoelectric relations such as Eqs.~\eqref{lorenz_bound},~\eqref{WF_generalized},~\eqref{thermal_hall_two}, and~\eqref{WF_sd_two} were derived. Some of them can be considered the reinterpretations of the classic Wiedemann-Franz law or the Mott's relation, and some other of them can help estimate how these many-body phases deviate from the free quasiparticle pictures. 

To extend this story, there are a couple of potentially fruitful open directions. Despite strong motivations from the solid-state systems, it is tempting to look for analogous relations in the context of holographic dualities, especially for dualities embracing electromagnetic dualities or the property of self-duality in the bulk-gravity theory such as the AdS${}_4$/CFT${}_3$ correspondence \cite{PhysRevD.75.085020, PhysRevD.76.106012, Melnikov2021}. We also look forward to generalizing the duality-based analyses to more bosonic systems and nonrelativistic inclined models. The latter predict measurable behaviors of experimentally accessible systems. The former opens up new routes to clarify natures of strongly correlated bosonic quantum matter lacking classical picture. For instance, the $\nu = 1$ fractional quantum Hall state can be dualized alternatively using a bilayer-graphenelike model with dynamical properties derivable by more conventional means \cite{PhysRevLett.117.136802, PhysRevResearch.3.013103}.  
\begin{acknowledgments}
We thank Yu-Ping Lin and Shih-An Wang for comments on the early version of this manuscript.
\end{acknowledgments}

\bibliography{citation}

\begin{thebibliography}{42}%
\makeatletter
\providecommand \@ifxundefined [1]{%
 \@ifx{#1\undefined}
}%
\providecommand \@ifnum [1]{%
 \ifnum #1\expandafter \@firstoftwo
 \else \expandafter \@secondoftwo
 \fi
}%
\providecommand \@ifx [1]{%
 \ifx #1\expandafter \@firstoftwo
 \else \expandafter \@secondoftwo
 \fi
}%
\providecommand \natexlab [1]{#1}%
\providecommand \enquote  [1]{``#1''}%
\providecommand \bibnamefont  [1]{#1}%
\providecommand \bibfnamefont [1]{#1}%
\providecommand \citenamefont [1]{#1}%
\providecommand \href@noop [0]{\@secondoftwo}%
\providecommand \href [0]{\begingroup \@sanitize@url \@href}%
\providecommand \@href[1]{\@@startlink{#1}\@@href}%
\providecommand \@@href[1]{\endgroup#1\@@endlink}%
\providecommand \@sanitize@url [0]{\catcode `\\12\catcode `\$12\catcode
  `\&12\catcode `\#12\catcode `\^12\catcode `\_12\catcode `\%12\relax}%
\providecommand \@@startlink[1]{}%
\providecommand \@@endlink[0]{}%
\providecommand \url  [0]{\begingroup\@sanitize@url \@url }%
\providecommand \@url [1]{\endgroup\@href {#1}{\urlprefix }}%
\providecommand \urlprefix  [0]{URL }%
\providecommand \Eprint [0]{\href }%
\providecommand \doibase [0]{https://doi.org/}%
\providecommand \selectlanguage [0]{\@gobble}%
\providecommand \bibinfo  [0]{\@secondoftwo}%
\providecommand \bibfield  [0]{\@secondoftwo}%
\providecommand \translation [1]{[#1]}%
\providecommand \BibitemOpen [0]{}%
\providecommand \bibitemStop [0]{}%
\providecommand \bibitemNoStop [0]{.\EOS\space}%
\providecommand \EOS [0]{\spacefactor3000\relax}%
\providecommand \BibitemShut  [1]{\csname bibitem#1\endcsname}%
\let\auto@bib@innerbib\@empty
\bibitem [{\citenamefont {Kane}\ and\ \citenamefont
  {Fisher}(1997)}]{PhysRevB.55.15832}%
  \BibitemOpen
  \bibfield  {author} {\bibinfo {author} {\bibfnamefont {C.~L.}\ \bibnamefont
  {Kane}}\ and\ \bibinfo {author} {\bibfnamefont {M.~P.~A.}\ \bibnamefont
  {Fisher}},\ }\bibfield  {title} {\bibinfo {title} {Quantized thermal
  transport in the fractional quantum hall effect},\ }\href
  {https://doi.org/10.1103/PhysRevB.55.15832} {\bibfield  {journal} {\bibinfo
  {journal} {Phys. Rev. B}\ }\textbf {\bibinfo {volume} {55}},\ \bibinfo
  {pages} {15832} (\bibinfo {year} {1997})}\BibitemShut {NoStop}%
\bibitem [{\citenamefont {Galitski}\ \emph {et~al.}(2005)\citenamefont
  {Galitski}, \citenamefont {Refael}, \citenamefont {Fisher},\ and\
  \citenamefont {Senthil}}]{PhysRevLett.95.077002}%
  \BibitemOpen
  \bibfield  {author} {\bibinfo {author} {\bibfnamefont {V.~M.}\ \bibnamefont
  {Galitski}}, \bibinfo {author} {\bibfnamefont {G.}~\bibnamefont {Refael}},
  \bibinfo {author} {\bibfnamefont {M.~P.~A.}\ \bibnamefont {Fisher}},\ and\
  \bibinfo {author} {\bibfnamefont {T.}~\bibnamefont {Senthil}},\ }\bibfield
  {title} {\bibinfo {title} {Vortices and quasiparticles near the
  superconductor-insulator transition in thin films},\ }\href
  {https://doi.org/10.1103/PhysRevLett.95.077002} {\bibfield  {journal}
  {\bibinfo  {journal} {Phys. Rev. Lett.}\ }\textbf {\bibinfo {volume} {95}},\
  \bibinfo {pages} {077002} (\bibinfo {year} {2005})}\BibitemShut {NoStop}%
\bibitem [{\citenamefont {Kim}\ and\ \citenamefont
  {P\'epin}(2009)}]{PhysRevLett.102.156404}%
  \BibitemOpen
  \bibfield  {author} {\bibinfo {author} {\bibfnamefont {K.-S.}\ \bibnamefont
  {Kim}}\ and\ \bibinfo {author} {\bibfnamefont {C.}~\bibnamefont {P\'epin}},\
  }\bibfield  {title} {\bibinfo {title} {Violation of the wiedemann-franz law
  at the kondo breakdown quantum critical point},\ }\href
  {https://doi.org/10.1103/PhysRevLett.102.156404} {\bibfield  {journal}
  {\bibinfo  {journal} {Phys. Rev. Lett.}\ }\textbf {\bibinfo {volume} {102}},\
  \bibinfo {pages} {156404} (\bibinfo {year} {2009})}\BibitemShut {NoStop}%
\bibitem [{\citenamefont {Principi}\ and\ \citenamefont
  {Vignale}(2015)}]{PhysRevLett.115.056603}%
  \BibitemOpen
  \bibfield  {author} {\bibinfo {author} {\bibfnamefont {A.}~\bibnamefont
  {Principi}}\ and\ \bibinfo {author} {\bibfnamefont {G.}~\bibnamefont
  {Vignale}},\ }\bibfield  {title} {\bibinfo {title} {Violation of the
  wiedemann-franz law in hydrodynamic electron liquids},\ }\href
  {https://doi.org/10.1103/PhysRevLett.115.056603} {\bibfield  {journal}
  {\bibinfo  {journal} {Phys. Rev. Lett.}\ }\textbf {\bibinfo {volume} {115}},\
  \bibinfo {pages} {056603} (\bibinfo {year} {2015})}\BibitemShut {NoStop}%
\bibitem [{\citenamefont {Lucas}\ and\ \citenamefont
  {Das~Sarma}(2018)}]{PhysRevB.97.245128}%
  \BibitemOpen
  \bibfield  {author} {\bibinfo {author} {\bibfnamefont {A.}~\bibnamefont
  {Lucas}}\ and\ \bibinfo {author} {\bibfnamefont {S.}~\bibnamefont
  {Das~Sarma}},\ }\bibfield  {title} {\bibinfo {title} {Electronic
  hydrodynamics and the breakdown of the wiedemann-franz and mott laws in
  interacting metals},\ }\href {https://doi.org/10.1103/PhysRevB.97.245128}
  {\bibfield  {journal} {\bibinfo  {journal} {Phys. Rev. B}\ }\textbf {\bibinfo
  {volume} {97}},\ \bibinfo {pages} {245128} (\bibinfo {year}
  {2018})}\BibitemShut {NoStop}%
\bibitem [{\citenamefont {Lavasani}\ \emph {et~al.}(2019)\citenamefont
  {Lavasani}, \citenamefont {Bulmash},\ and\ \citenamefont
  {Das~Sarma}}]{PhysRevB.99.085104}%
  \BibitemOpen
  \bibfield  {author} {\bibinfo {author} {\bibfnamefont {A.}~\bibnamefont
  {Lavasani}}, \bibinfo {author} {\bibfnamefont {D.}~\bibnamefont {Bulmash}},\
  and\ \bibinfo {author} {\bibfnamefont {S.}~\bibnamefont {Das~Sarma}},\
  }\bibfield  {title} {\bibinfo {title} {Wiedemann-franz law and fermi
  liquids},\ }\href {https://doi.org/10.1103/PhysRevB.99.085104} {\bibfield
  {journal} {\bibinfo  {journal} {Phys. Rev. B}\ }\textbf {\bibinfo {volume}
  {99}},\ \bibinfo {pages} {085104} (\bibinfo {year} {2019})}\BibitemShut
  {NoStop}%
\bibitem [{\citenamefont {Melnikov}\ and\ \citenamefont
  {Nastase}(2021)}]{Melnikov2021}%
  \BibitemOpen
  \bibfield  {author} {\bibinfo {author} {\bibfnamefont {D.}~\bibnamefont
  {Melnikov}}\ and\ \bibinfo {author} {\bibfnamefont {H.}~\bibnamefont
  {Nastase}},\ }\bibfield  {title} {\bibinfo {title} {Wiedemann-franz laws and
  sl(2,z) duality in ads/cmt holographic duals and one-dimensional effective
  actions for them},\ }\href {https://doi.org/10.1007/JHEP05(2021)092}
  {\bibfield  {journal} {\bibinfo  {journal} {J. High Energy Phys.}\ }\textbf
  {\bibinfo {volume} {2021}}\bibinfo  {number} { (5)},\ \bibinfo {pages}
  {92}}\BibitemShut {NoStop}%
\bibitem [{Note1()}]{Note1}%
  \BibitemOpen
\bibfield  {number} {  }\bibinfo {note} {The parity anomaly would not concern
  us since we have coupled the Dirac cone to a (3+1)-dimensional field \cite
  {PhysRevB.88.085104}.}\BibitemShut {Stop}%
\bibitem [{\citenamefont {Hsiao}\ and\ \citenamefont
  {Son}(2017)}]{PhysRevB.96.075127}%
  \BibitemOpen
  \bibfield  {author} {\bibinfo {author} {\bibfnamefont {W.-H.}\ \bibnamefont
  {Hsiao}}\ and\ \bibinfo {author} {\bibfnamefont {D.~T.}\ \bibnamefont
  {Son}},\ }\bibfield  {title} {\bibinfo {title} {Duality and universal
  transport in mixed-dimension electrodynamics},\ }\href
  {https://doi.org/10.1103/PhysRevB.96.075127} {\bibfield  {journal} {\bibinfo
  {journal} {Phys. Rev. B}\ }\textbf {\bibinfo {volume} {96}},\ \bibinfo
  {pages} {075127} (\bibinfo {year} {2017})}\BibitemShut {NoStop}%
\bibitem [{\citenamefont {Hsiao}\ and\ \citenamefont
  {Son}(2019)}]{PhysRevB.100.235150}%
  \BibitemOpen
  \bibfield  {author} {\bibinfo {author} {\bibfnamefont {W.-H.}\ \bibnamefont
  {Hsiao}}\ and\ \bibinfo {author} {\bibfnamefont {D.~T.}\ \bibnamefont
  {Son}},\ }\bibfield  {title} {\bibinfo {title} {Self-dual
  $\ensuremath{\nu}=1$ bosonic quantum hall state in mixed-dimensional qed},\
  }\href {https://doi.org/10.1103/PhysRevB.100.235150} {\bibfield  {journal}
  {\bibinfo  {journal} {Phys. Rev. B}\ }\textbf {\bibinfo {volume} {100}},\
  \bibinfo {pages} {235150} (\bibinfo {year} {2019})}\BibitemShut {NoStop}%
\bibitem [{\citenamefont {Peskin}(1978)}]{PESKIN1978122}%
  \BibitemOpen
  \bibfield  {author} {\bibinfo {author} {\bibfnamefont {M.~E.}\ \bibnamefont
  {Peskin}},\ }\bibfield  {title} {\bibinfo {title} {Mandelstam-'t hooft
  duality in abelian lattice models},\ }\href
  {https://doi.org/https://doi.org/10.1016/0003-4916(78)90252-X} {\bibfield
  {journal} {\bibinfo  {journal} {Annals of Physics}\ }\textbf {\bibinfo
  {volume} {113}},\ \bibinfo {pages} {122} (\bibinfo {year}
  {1978})}\BibitemShut {NoStop}%
\bibitem [{\citenamefont {Dasgupta}\ and\ \citenamefont
  {Halperin}(1981)}]{PhysRevLett.47.1556}%
  \BibitemOpen
  \bibfield  {author} {\bibinfo {author} {\bibfnamefont {C.}~\bibnamefont
  {Dasgupta}}\ and\ \bibinfo {author} {\bibfnamefont {B.~I.}\ \bibnamefont
  {Halperin}},\ }\bibfield  {title} {\bibinfo {title} {Phase transition in a
  lattice model of superconductivity},\ }\href
  {https://doi.org/10.1103/PhysRevLett.47.1556} {\bibfield  {journal} {\bibinfo
   {journal} {Phys. Rev. Lett.}\ }\textbf {\bibinfo {volume} {47}},\ \bibinfo
  {pages} {1556} (\bibinfo {year} {1981})}\BibitemShut {NoStop}%
\bibitem [{\citenamefont {Seiberg}\ \emph {et~al.}(2016)\citenamefont
  {Seiberg}, \citenamefont {Senthil}, \citenamefont {Wang},\ and\ \citenamefont
  {Witten}}]{SEIBERG2016395}%
  \BibitemOpen
  \bibfield  {author} {\bibinfo {author} {\bibfnamefont {N.}~\bibnamefont
  {Seiberg}}, \bibinfo {author} {\bibfnamefont {T.}~\bibnamefont {Senthil}},
  \bibinfo {author} {\bibfnamefont {C.}~\bibnamefont {Wang}},\ and\ \bibinfo
  {author} {\bibfnamefont {E.}~\bibnamefont {Witten}},\ }\bibfield  {title}
  {\bibinfo {title} {A duality web in 2+1 dimensions and condensed matter
  physics},\ }\href {https://doi.org/https://doi.org/10.1016/j.aop.2016.08.007}
  {\bibfield  {journal} {\bibinfo  {journal} {Annals of Physics}\ }\textbf
  {\bibinfo {volume} {374}},\ \bibinfo {pages} {395 } (\bibinfo {year}
  {2016})}\BibitemShut {NoStop}%
\bibitem [{\citenamefont {Karch}\ and\ \citenamefont
  {Tong}(2016)}]{PhysRevX.6.031043}%
  \BibitemOpen
  \bibfield  {author} {\bibinfo {author} {\bibfnamefont {A.}~\bibnamefont
  {Karch}}\ and\ \bibinfo {author} {\bibfnamefont {D.}~\bibnamefont {Tong}},\
  }\bibfield  {title} {\bibinfo {title} {Particle-vortex duality from 3d
  bosonization},\ }\href {https://doi.org/10.1103/PhysRevX.6.031043} {\bibfield
   {journal} {\bibinfo  {journal} {Phys. Rev. X}\ }\textbf {\bibinfo {volume}
  {6}},\ \bibinfo {pages} {031043} (\bibinfo {year} {2016})}\BibitemShut
  {NoStop}%
\bibitem [{\citenamefont {Teber}(2012)}]{PhysRevD.86.025005}%
  \BibitemOpen
  \bibfield  {author} {\bibinfo {author} {\bibfnamefont {S.}~\bibnamefont
  {Teber}},\ }\bibfield  {title} {\bibinfo {title} {Electromagnetic current
  correlations in reduced quantum electrodynamics},\ }\href
  {https://doi.org/10.1103/PhysRevD.86.025005} {\bibfield  {journal} {\bibinfo
  {journal} {Phys. Rev. D}\ }\textbf {\bibinfo {volume} {86}},\ \bibinfo
  {pages} {025005} (\bibinfo {year} {2012})}\BibitemShut {NoStop}%
\bibitem [{\citenamefont {Teber}\ and\ \citenamefont
  {Kotikov}(2018)}]{PhysRevD.97.074004}%
  \BibitemOpen
  \bibfield  {author} {\bibinfo {author} {\bibfnamefont {S.}~\bibnamefont
  {Teber}}\ and\ \bibinfo {author} {\bibfnamefont {A.~V.}\ \bibnamefont
  {Kotikov}},\ }\bibfield  {title} {\bibinfo {title} {Field theoretic
  renormalization study of reduced quantum electrodynamics and applications to
  the ultrarelativistic limit of dirac liquids},\ }\href
  {https://doi.org/10.1103/PhysRevD.97.074004} {\bibfield  {journal} {\bibinfo
  {journal} {Phys. Rev. D}\ }\textbf {\bibinfo {volume} {97}},\ \bibinfo
  {pages} {074004} (\bibinfo {year} {2018})}\BibitemShut {NoStop}%
\bibitem [{\citenamefont {Dudal}\ \emph {et~al.}(2019)\citenamefont {Dudal},
  \citenamefont {Mizher},\ and\ \citenamefont {Pais}}]{PhysRevD.99.045017}%
  \BibitemOpen
  \bibfield  {author} {\bibinfo {author} {\bibfnamefont {D.}~\bibnamefont
  {Dudal}}, \bibinfo {author} {\bibfnamefont {A.~J.}\ \bibnamefont {Mizher}},\
  and\ \bibinfo {author} {\bibfnamefont {P.}~\bibnamefont {Pais}},\ }\bibfield
  {title} {\bibinfo {title} {Exact quantum scale invariance of
  three-dimensional reduced qed theories},\ }\href
  {https://doi.org/10.1103/PhysRevD.99.045017} {\bibfield  {journal} {\bibinfo
  {journal} {Phys. Rev. D}\ }\textbf {\bibinfo {volume} {99}},\ \bibinfo
  {pages} {045017} (\bibinfo {year} {2019})}\BibitemShut {NoStop}%
\bibitem [{\citenamefont {Son}(2015)}]{PhysRevX.5.031027}%
  \BibitemOpen
  \bibfield  {author} {\bibinfo {author} {\bibfnamefont {D.~T.}\ \bibnamefont
  {Son}},\ }\bibfield  {title} {\bibinfo {title} {Is the composite fermion a
  dirac particle?},\ }\href {https://doi.org/10.1103/PhysRevX.5.031027}
  {\bibfield  {journal} {\bibinfo  {journal} {Phys. Rev. X}\ }\textbf {\bibinfo
  {volume} {5}},\ \bibinfo {pages} {031027} (\bibinfo {year}
  {2015})}\BibitemShut {NoStop}%
\bibitem [{\citenamefont {Wang}\ and\ \citenamefont
  {Senthil}(2015)}]{PhysRevX.5.041031}%
  \BibitemOpen
  \bibfield  {author} {\bibinfo {author} {\bibfnamefont {C.}~\bibnamefont
  {Wang}}\ and\ \bibinfo {author} {\bibfnamefont {T.}~\bibnamefont {Senthil}},\
  }\bibfield  {title} {\bibinfo {title} {Dual dirac liquid on the surface of
  the electron topological insulator},\ }\href
  {https://doi.org/10.1103/PhysRevX.5.041031} {\bibfield  {journal} {\bibinfo
  {journal} {Phys. Rev. X}\ }\textbf {\bibinfo {volume} {5}},\ \bibinfo {pages}
  {041031} (\bibinfo {year} {2015})}\BibitemShut {NoStop}%
\bibitem [{\citenamefont {Metlitski}\ and\ \citenamefont
  {Vishwanath}(2016)}]{PhysRevB.93.245151}%
  \BibitemOpen
  \bibfield  {author} {\bibinfo {author} {\bibfnamefont {M.~A.}\ \bibnamefont
  {Metlitski}}\ and\ \bibinfo {author} {\bibfnamefont {A.}~\bibnamefont
  {Vishwanath}},\ }\bibfield  {title} {\bibinfo {title} {Particle-vortex
  duality of two-dimensional dirac fermion from electric-magnetic duality of
  three-dimensional topological insulators},\ }\href
  {https://doi.org/10.1103/PhysRevB.93.245151} {\bibfield  {journal} {\bibinfo
  {journal} {Phys. Rev. B}\ }\textbf {\bibinfo {volume} {93}},\ \bibinfo
  {pages} {245151} (\bibinfo {year} {2016})}\BibitemShut {NoStop}%
\bibitem [{Note2()}]{Note2}%
  \BibitemOpen
  \bibinfo {note} {To avoid confusion, in this work we adopt the convention
  $A_{\mu } = (A^0, -\protect \mathbf A)$ for U(1) gauge fields. Therefore,
  $\epsilon ^{ij}\partial _iA_j = -B$ and $-\partial _0A_i - \partial _i A_0 =
  E_i$. It is straightforward to verify this convention consistently yields all
  electromagnetism.}\BibitemShut {Stop}%
\bibitem [{\citenamefont {Wen}\ and\ \citenamefont
  {Zee}(1990)}]{doi:10.1142/S0217979290000206}%
  \BibitemOpen
  \bibfield  {author} {\bibinfo {author} {\bibfnamefont {X.~G.}\ \bibnamefont
  {Wen}}\ and\ \bibinfo {author} {\bibfnamefont {A.}~\bibnamefont {Zee}},\
  }\bibfield  {title} {\bibinfo {title} {Universal conductance at the
  superconductor-insulator transition},\ }\href
  {https://doi.org/10.1142/S0217979290000206} {\bibfield  {journal} {\bibinfo
  {journal} {Int. J. Mod. Phys. B}\ }\textbf {\bibinfo {volume} {04}},\
  \bibinfo {pages} {437} (\bibinfo {year} {1990})}\BibitemShut {NoStop}%
\bibitem [{\citenamefont {Burgess}\ and\ \citenamefont
  {Dolan}(2001)}]{PhysRevB.63.155309}%
  \BibitemOpen
  \bibfield  {author} {\bibinfo {author} {\bibfnamefont {C.~P.}\ \bibnamefont
  {Burgess}}\ and\ \bibinfo {author} {\bibfnamefont {B.~P.}\ \bibnamefont
  {Dolan}},\ }\bibfield  {title} {\bibinfo {title} {Particle-vortex duality and
  the modular group: Applications to the quantum hall effect and other
  two-dimensional systems},\ }\href
  {https://doi.org/10.1103/PhysRevB.63.155309} {\bibfield  {journal} {\bibinfo
  {journal} {Phys. Rev. B}\ }\textbf {\bibinfo {volume} {63}},\ \bibinfo
  {pages} {155309} (\bibinfo {year} {2001})}\BibitemShut {NoStop}%
\bibitem [{\citenamefont {Hartnoll}\ \emph {et~al.}(2007)\citenamefont
  {Hartnoll}, \citenamefont {Kovtun}, \citenamefont {M\"uller},\ and\
  \citenamefont {Sachdev}}]{PhysRevB.76.144502}%
  \BibitemOpen
  \bibfield  {author} {\bibinfo {author} {\bibfnamefont {S.~A.}\ \bibnamefont
  {Hartnoll}}, \bibinfo {author} {\bibfnamefont {P.~K.}\ \bibnamefont
  {Kovtun}}, \bibinfo {author} {\bibfnamefont {M.}~\bibnamefont {M\"uller}},\
  and\ \bibinfo {author} {\bibfnamefont {S.}~\bibnamefont {Sachdev}},\
  }\bibfield  {title} {\bibinfo {title} {Theory of the nernst effect near
  quantum phase transitions in condensed matter and in dyonic black holes},\
  }\href {https://doi.org/10.1103/PhysRevB.76.144502} {\bibfield  {journal}
  {\bibinfo  {journal} {Phys. Rev. B}\ }\textbf {\bibinfo {volume} {76}},\
  \bibinfo {pages} {144502} (\bibinfo {year} {2007})}\BibitemShut {NoStop}%
\bibitem [{\citenamefont {Donos}\ \emph {et~al.}(2017)\citenamefont {Donos},
  \citenamefont {Gauntlett}, \citenamefont {Griffin}, \citenamefont
  {Lohitsiri},\ and\ \citenamefont {Melgar}}]{Donos2017}%
  \BibitemOpen
  \bibfield  {author} {\bibinfo {author} {\bibfnamefont {A.}~\bibnamefont
  {Donos}}, \bibinfo {author} {\bibfnamefont {J.~P.}\ \bibnamefont
  {Gauntlett}}, \bibinfo {author} {\bibfnamefont {T.}~\bibnamefont {Griffin}},
  \bibinfo {author} {\bibfnamefont {N.}~\bibnamefont {Lohitsiri}},\ and\
  \bibinfo {author} {\bibfnamefont {L.}~\bibnamefont {Melgar}},\ }\bibfield
  {title} {\bibinfo {title} {Holographic dc conductivity and onsager
  relations},\ }\href {https://doi.org/10.1007/JHEP07(2017)006} {\bibfield
  {journal} {\bibinfo  {journal} {J. High Energy Phys.}\ }\textbf {\bibinfo
  {volume} {07}},\ \bibinfo {pages} {006}}\BibitemShut {NoStop}%
\bibitem [{\citenamefont {Hartnoll}\ and\ \citenamefont
  {Herzog}(2007)}]{PhysRevD.76.106012}%
  \BibitemOpen
  \bibfield  {author} {\bibinfo {author} {\bibfnamefont {S.~A.}\ \bibnamefont
  {Hartnoll}}\ and\ \bibinfo {author} {\bibfnamefont {C.~P.}\ \bibnamefont
  {Herzog}},\ }\bibfield  {title} {\bibinfo {title} {Ohm's law at strong
  coupling: S duality and the cyclotron resonance},\ }\href
  {https://doi.org/10.1103/PhysRevD.76.106012} {\bibfield  {journal} {\bibinfo
  {journal} {Phys. Rev. D}\ }\textbf {\bibinfo {volume} {76}},\ \bibinfo
  {pages} {106012} (\bibinfo {year} {2007})}\BibitemShut {NoStop}%
\bibitem [{\citenamefont {Crossno}\ \emph {et~al.}(2016)\citenamefont
  {Crossno}, \citenamefont {Shi}, \citenamefont {Wang}, \citenamefont {Liu},
  \citenamefont {Harzheim}, \citenamefont {Lucas}, \citenamefont {Sachdev},
  \citenamefont {Kim}, \citenamefont {Taniguchi}, \citenamefont {Watanabe},
  \citenamefont {Ohki},\ and\ \citenamefont {Fong}}]{Crossno1058}%
  \BibitemOpen
  \bibfield  {author} {\bibinfo {author} {\bibfnamefont {J.}~\bibnamefont
  {Crossno}}, \bibinfo {author} {\bibfnamefont {J.~K.}\ \bibnamefont {Shi}},
  \bibinfo {author} {\bibfnamefont {K.}~\bibnamefont {Wang}}, \bibinfo {author}
  {\bibfnamefont {X.}~\bibnamefont {Liu}}, \bibinfo {author} {\bibfnamefont
  {A.}~\bibnamefont {Harzheim}}, \bibinfo {author} {\bibfnamefont
  {A.}~\bibnamefont {Lucas}}, \bibinfo {author} {\bibfnamefont
  {S.}~\bibnamefont {Sachdev}}, \bibinfo {author} {\bibfnamefont
  {P.}~\bibnamefont {Kim}}, \bibinfo {author} {\bibfnamefont {T.}~\bibnamefont
  {Taniguchi}}, \bibinfo {author} {\bibfnamefont {K.}~\bibnamefont {Watanabe}},
  \bibinfo {author} {\bibfnamefont {T.~A.}\ \bibnamefont {Ohki}},\ and\
  \bibinfo {author} {\bibfnamefont {K.~C.}\ \bibnamefont {Fong}},\ }\bibfield
  {title} {\bibinfo {title} {Observation of the dirac fluid and the breakdown
  of the wiedemann-franz law in graphene},\ }\href
  {https://doi.org/10.1126/science.aad0343} {\bibfield  {journal} {\bibinfo
  {journal} {Science}\ }\textbf {\bibinfo {volume} {351}},\ \bibinfo {pages}
  {1058} (\bibinfo {year} {2016})}\BibitemShut {NoStop}%
\bibitem [{\citenamefont {Lucas}\ \emph {et~al.}(2016)\citenamefont {Lucas},
  \citenamefont {Crossno}, \citenamefont {Fong}, \citenamefont {Kim},\ and\
  \citenamefont {Sachdev}}]{PhysRevB.93.075426}%
  \BibitemOpen
  \bibfield  {author} {\bibinfo {author} {\bibfnamefont {A.}~\bibnamefont
  {Lucas}}, \bibinfo {author} {\bibfnamefont {J.}~\bibnamefont {Crossno}},
  \bibinfo {author} {\bibfnamefont {K.~C.}\ \bibnamefont {Fong}}, \bibinfo
  {author} {\bibfnamefont {P.}~\bibnamefont {Kim}},\ and\ \bibinfo {author}
  {\bibfnamefont {S.}~\bibnamefont {Sachdev}},\ }\bibfield  {title} {\bibinfo
  {title} {Transport in inhomogeneous quantum critical fluids and in the dirac
  fluid in graphene},\ }\href {https://doi.org/10.1103/PhysRevB.93.075426}
  {\bibfield  {journal} {\bibinfo  {journal} {Phys. Rev. B}\ }\textbf {\bibinfo
  {volume} {93}},\ \bibinfo {pages} {075426} (\bibinfo {year}
  {2016})}\BibitemShut {NoStop}%
\bibitem [{\citenamefont {Sharapov}\ \emph {et~al.}(2003)\citenamefont
  {Sharapov}, \citenamefont {Gusynin},\ and\ \citenamefont
  {Beck}}]{PhysRevB.67.144509}%
  \BibitemOpen
  \bibfield  {author} {\bibinfo {author} {\bibfnamefont {S.~G.}\ \bibnamefont
  {Sharapov}}, \bibinfo {author} {\bibfnamefont {V.~P.}\ \bibnamefont
  {Gusynin}},\ and\ \bibinfo {author} {\bibfnamefont {H.}~\bibnamefont
  {Beck}},\ }\bibfield  {title} {\bibinfo {title} {Transport properties in the
  d-density-wave state in an external magnetic field: The wiedemann-franz
  law},\ }\href {https://doi.org/10.1103/PhysRevB.67.144509} {\bibfield
  {journal} {\bibinfo  {journal} {Phys. Rev. B}\ }\textbf {\bibinfo {volume}
  {67}},\ \bibinfo {pages} {144509} (\bibinfo {year} {2003})}\BibitemShut
  {NoStop}%
\bibitem [{\citenamefont {Rycerz}(2021)}]{ma14112704}%
  \BibitemOpen
  \bibfield  {author} {\bibinfo {author} {\bibfnamefont {A.}~\bibnamefont
  {Rycerz}},\ }\bibfield  {title} {\bibinfo {title} {Wiedemann–franz law for
  massless dirac fermions with implications for graphene},\ }\href
  {https://www.mdpi.com/1996-1944/14/11/2704} {\bibfield  {journal} {\bibinfo
  {journal} {Materials}\ }\textbf {\bibinfo {volume} {14}} (\bibinfo {year}
  {2021})}\BibitemShut {NoStop}%
\bibitem [{Note3()}]{Note3}%
  \BibitemOpen
  \bibinfo {note} {Note that the dimensionless $\sigma $ measures the
  coefficient of $e^2/\hbar $ for the empirical conductivity. Therefore, the
  $\rho $ in this equation should also be understood as the coefficient of
  $\hbar /e^2$ for the empirical resistivity in the corresponding
  system.}\BibitemShut {Stop}%
\bibitem [{\citenamefont {Wang}\ and\ \citenamefont
  {Senthil}(2016)}]{PhysRevB.93.085110}%
  \BibitemOpen
  \bibfield  {author} {\bibinfo {author} {\bibfnamefont {C.}~\bibnamefont
  {Wang}}\ and\ \bibinfo {author} {\bibfnamefont {T.}~\bibnamefont {Senthil}},\
  }\bibfield  {title} {\bibinfo {title} {Half-filled landau level, topological
  insulator surfaces, and three-dimensional quantum spin liquids},\ }\href
  {https://doi.org/10.1103/PhysRevB.93.085110} {\bibfield  {journal} {\bibinfo
  {journal} {Phys. Rev. B}\ }\textbf {\bibinfo {volume} {93}},\ \bibinfo
  {pages} {085110} (\bibinfo {year} {2016})}\BibitemShut {NoStop}%
\bibitem [{\citenamefont {Potter}\ \emph {et~al.}(2016)\citenamefont {Potter},
  \citenamefont {Serbyn},\ and\ \citenamefont
  {Vishwanath}}]{PhysRevX.6.031026}%
  \BibitemOpen
  \bibfield  {author} {\bibinfo {author} {\bibfnamefont {A.~C.}\ \bibnamefont
  {Potter}}, \bibinfo {author} {\bibfnamefont {M.}~\bibnamefont {Serbyn}},\
  and\ \bibinfo {author} {\bibfnamefont {A.}~\bibnamefont {Vishwanath}},\
  }\bibfield  {title} {\bibinfo {title} {Thermoelectric transport signatures of
  dirac composite fermions in the half-filled landau level},\ }\href
  {https://doi.org/10.1103/PhysRevX.6.031026} {\bibfield  {journal} {\bibinfo
  {journal} {Phys. Rev. X}\ }\textbf {\bibinfo {volume} {6}},\ \bibinfo {pages}
  {031026} (\bibinfo {year} {2016})}\BibitemShut {NoStop}%
\bibitem [{\citenamefont {Nguyen}\ and\ \citenamefont
  {Gromov}(2017)}]{PhysRevB.95.085151}%
  \BibitemOpen
  \bibfield  {author} {\bibinfo {author} {\bibfnamefont {D.~X.}\ \bibnamefont
  {Nguyen}}\ and\ \bibinfo {author} {\bibfnamefont {A.}~\bibnamefont
  {Gromov}},\ }\bibfield  {title} {\bibinfo {title} {Exact electromagnetic
  response of landau level electrons},\ }\href
  {https://doi.org/10.1103/PhysRevB.95.085151} {\bibfield  {journal} {\bibinfo
  {journal} {Phys. Rev. B}\ }\textbf {\bibinfo {volume} {95}},\ \bibinfo
  {pages} {085151} (\bibinfo {year} {2017})}\BibitemShut {NoStop}%
\bibitem [{\citenamefont {Mulligan}\ and\ \citenamefont
  {Raghu}(2016)}]{PhysRevB.93.205116}%
  \BibitemOpen
  \bibfield  {author} {\bibinfo {author} {\bibfnamefont {M.}~\bibnamefont
  {Mulligan}}\ and\ \bibinfo {author} {\bibfnamefont {S.}~\bibnamefont
  {Raghu}},\ }\bibfield  {title} {\bibinfo {title} {Composite fermions and the
  field-tuned superconductor-insulator transition},\ }\href
  {https://doi.org/10.1103/PhysRevB.93.205116} {\bibfield  {journal} {\bibinfo
  {journal} {Phys. Rev. B}\ }\textbf {\bibinfo {volume} {93}},\ \bibinfo
  {pages} {205116} (\bibinfo {year} {2016})}\BibitemShut {NoStop}%
\bibitem [{\citenamefont {Mulligan}(2017)}]{PhysRevB.95.045118}%
  \BibitemOpen
  \bibfield  {author} {\bibinfo {author} {\bibfnamefont {M.}~\bibnamefont
  {Mulligan}},\ }\bibfield  {title} {\bibinfo {title} {Particle-vortex
  symmetric liquid},\ }\href {https://doi.org/10.1103/PhysRevB.95.045118}
  {\bibfield  {journal} {\bibinfo  {journal} {Phys. Rev. B}\ }\textbf {\bibinfo
  {volume} {95}},\ \bibinfo {pages} {045118} (\bibinfo {year}
  {2017})}\BibitemShut {NoStop}%
\bibitem [{\citenamefont {Pasquier}\ and\ \citenamefont
  {Haldane}(1998)}]{PASQUIER1998719}%
  \BibitemOpen
  \bibfield  {author} {\bibinfo {author} {\bibfnamefont {V.}~\bibnamefont
  {Pasquier}}\ and\ \bibinfo {author} {\bibfnamefont {F.}~\bibnamefont
  {Haldane}},\ }\bibfield  {title} {\bibinfo {title} {A dipole interpretation
  of the $\nu = 1/2$ state},\ }\href
  {https://doi.org/https://doi.org/10.1016/S0550-3213(98)00069-8} {\bibfield
  {journal} {\bibinfo  {journal} {Nucl. Phys. B}\ }\textbf {\bibinfo {volume}
  {516}},\ \bibinfo {pages} {719 } (\bibinfo {year} {1998})}\BibitemShut
  {NoStop}%
\bibitem [{\citenamefont {Read}(1998)}]{PhysRevB.58.16262}%
  \BibitemOpen
  \bibfield  {author} {\bibinfo {author} {\bibfnamefont {N.}~\bibnamefont
  {Read}},\ }\bibfield  {title} {\bibinfo {title} {Lowest-landau-level theory
  of the quantum hall effect: The fermi-liquid-like state of bosons at filling
  factor one},\ }\href {https://doi.org/10.1103/PhysRevB.58.16262} {\bibfield
  {journal} {\bibinfo  {journal} {Phys. Rev. B}\ }\textbf {\bibinfo {volume}
  {58}},\ \bibinfo {pages} {16262} (\bibinfo {year} {1998})}\BibitemShut
  {NoStop}%
\bibitem [{\citenamefont {Herzog}\ \emph {et~al.}(2007)\citenamefont {Herzog},
  \citenamefont {Kovtun}, \citenamefont {Sachdev},\ and\ \citenamefont
  {Son}}]{PhysRevD.75.085020}%
  \BibitemOpen
  \bibfield  {author} {\bibinfo {author} {\bibfnamefont {C.~P.}\ \bibnamefont
  {Herzog}}, \bibinfo {author} {\bibfnamefont {P.}~\bibnamefont {Kovtun}},
  \bibinfo {author} {\bibfnamefont {S.}~\bibnamefont {Sachdev}},\ and\ \bibinfo
  {author} {\bibfnamefont {D.~T.}\ \bibnamefont {Son}},\ }\bibfield  {title}
  {\bibinfo {title} {Quantum critical transport, duality, and m theory},\
  }\href {https://doi.org/10.1103/PhysRevD.75.085020} {\bibfield  {journal}
  {\bibinfo  {journal} {Phys. Rev. D}\ }\textbf {\bibinfo {volume} {75}},\
  \bibinfo {pages} {085020} (\bibinfo {year} {2007})}\BibitemShut {NoStop}%
\bibitem [{\citenamefont {Mross}\ \emph {et~al.}(2016)\citenamefont {Mross},
  \citenamefont {Alicea},\ and\ \citenamefont
  {Motrunich}}]{PhysRevLett.117.136802}%
  \BibitemOpen
  \bibfield  {author} {\bibinfo {author} {\bibfnamefont {D.~F.}\ \bibnamefont
  {Mross}}, \bibinfo {author} {\bibfnamefont {J.}~\bibnamefont {Alicea}},\ and\
  \bibinfo {author} {\bibfnamefont {O.~I.}\ \bibnamefont {Motrunich}},\
  }\bibfield  {title} {\bibinfo {title} {Bosonic analogue of dirac composite
  fermi liquid},\ }\href {https://doi.org/10.1103/PhysRevLett.117.136802}
  {\bibfield  {journal} {\bibinfo  {journal} {Phys. Rev. Lett.}\ }\textbf
  {\bibinfo {volume} {117}},\ \bibinfo {pages} {136802} (\bibinfo {year}
  {2016})}\BibitemShut {NoStop}%
\bibitem [{\citenamefont {Hsiao}(2021)}]{PhysRevResearch.3.013103}%
  \BibitemOpen
  \bibfield  {author} {\bibinfo {author} {\bibfnamefont {W.-H.}\ \bibnamefont
  {Hsiao}},\ }\bibfield  {title} {\bibinfo {title} {Time-reversal odd transport
  in bilayer graphene: Hall conductivity and hall viscosity},\ }\href
  {https://doi.org/10.1103/PhysRevResearch.3.013103} {\bibfield  {journal}
  {\bibinfo  {journal} {Phys. Rev. Research}\ }\textbf {\bibinfo {volume}
  {3}},\ \bibinfo {pages} {013103} (\bibinfo {year} {2021})}\BibitemShut
  {NoStop}%
\bibitem [{\citenamefont {Mulligan}\ and\ \citenamefont
  {Burnell}(2013)}]{PhysRevB.88.085104}%
  \BibitemOpen
  \bibfield  {author} {\bibinfo {author} {\bibfnamefont {M.}~\bibnamefont
  {Mulligan}}\ and\ \bibinfo {author} {\bibfnamefont {F.~J.}\ \bibnamefont
  {Burnell}},\ }\bibfield  {title} {\bibinfo {title} {Topological insulators
  avoid the parity anomaly},\ }\href
  {https://doi.org/10.1103/PhysRevB.88.085104} {\bibfield  {journal} {\bibinfo
  {journal} {Phys. Rev. B}\ }\textbf {\bibinfo {volume} {88}},\ \bibinfo
  {pages} {085104} (\bibinfo {year} {2013})}\BibitemShut {NoStop}%
\end{thebibliography}%
\end{document}